# Machine learning-based classification of variable stars using phase-folded light curves

Almat Akhmetali[a], Alisher Zhunuskanov[a], Timur Namazbayev[a], Marat Zaidyn[a],
Aknur Sakan[a], Dana Turlykozhayeva[a], Nurzhan Ussipov[a]

[a]Department of Electronics and Astrophysics, Al-Farabi Kazakh National University, 050040 Almaty, Kazakhstan

## ABSTRACT

Classifying variable stars is crucial for advancing our understanding of stellar evolution and dynamics. As large-scale surveys generate increasing volumes of light curve data, the demand for automated and reliable classification techniques continues to grow. Traditional methods often rely on manual feature extraction and selection, which can be labor-intensive and less effective for managing extensive datasets.

In this study, we present a convolutional neural network (CNN)-based method for classifying variable stars using raw light curve data and their known periods. Our approach eliminates the need for manual feature extraction and preselected preprocessing steps. By applying phase-folding and interpolation to structure the light curves, the model learns variability patterns critical for accurate classification.

Trained and evaluated on the All-Sky Automated Survey for Supernovae (ASAS-SN) dataset, our model achieves an average accuracy of 90% and an F1 score of 0.86 across six well-known classes of variable stars. The CNN effectively handles the diverse shapes and sampling cadences of light curves, offering a robust, automated, data-driven solution for classifying variable stars. This automated, data-driven method provides a robust solution for classifying variable stars, enabling the efficient analysis of large datasets from both current and future sky surveys.

## 1. INTRODUCTION

Variable stars are fundamental to enhancing our understanding of the universe, playing a pivotal role in areas such as stellar and galactic astrophysics as well as cosmology. These stars have been key to groundbreaking discoveries, including calculating distances to galaxies, determining the Hubble constant, exploring stellar evolution, investigating planetary formation, and analyzing the chemical composition of various galactic regions [1–5]. The effective study of variable stars has been greatly facilitated by modern astronomical surveys, which have produced vast amounts of data. The development of advanced astronomical instruments and large-scale time-domain surveys, such as the All-Sky Automated Survey for Supernovae (ASAS-SN) [6], the Optical Gravitational Lensing Experiment (OGLE) [7], the Zwicky Transient Facility (ZTF) [8], the Catalina Real-Time Transient Survey (CRTS) [9] and NASA's Kepler mission [10] has led to an exponential increase in the availability of time-series data. These datasets, which consist of light curves representing the brightness variations of stars over time, are invaluable for analyzing variable stars. However, the sheer scale and complexity of these data necessitate the use of automated classification techniques. Automatic classification not only addresses the impracticality of manual methods for handling such vast datasets but also ensures that astronomers can extract maximum scientific insights from these surveys.

Traditional methods for classifying variable stars typically involve deriving features from light curves, such as statistical metrics, period analyses, and Fourier decomposition parameters [11]. These features distill the information from each light curve into concise representations for machine learning (ML) classification. Although these methods can be effective, they are often computationally intensive and specifically designed for particular survey data, reducing their flexibility. Additionally, such approaches face challenges when dealing with light curves that are sparse, noisy, or irregularly sampled due to observational limitations, complicating the classification process and diminishing reliability across diverse datasets. For example, Debosscher et al. [12] developed an automated method that utilized 28 features derived from the Fourier analysis of time-series data, focusing on amplitudes, phases, and frequencies obtained through Fourier fitting. These features were then input into Gaussian Mixture and ML classifiers for supervised learning. Similarly, Kim et al. [13] concentrated on detecting Quasi-Stellar Objects (QSOs) within the MACHO dataset [14], finding that the Random Forest (RF) classifier [15] outperformed the Support Vector Machine (SVM) [16,17] when using 11 features. In the same year,

Richards et al. [18] achieved similar results by evaluating 53 features, combining periodic features with non-periodic ones proposed by Butler and Bloom [19]. They demonstrated the application of various ML-based classifiers for the automatic classification of numerous variability classes. Their work also explored hierarchical classification approaches, employing hierarchical single-label classification (HSC) and hierarchical multi-label classification (HMC) with RFs. Additionally, Kim and Bailer-Jones developed the UPSILoN package, which extracts 16 features from light curves and classifies them using the RF technique [20]. More recent studies have continued to utilize these methods to identify specific types of variability [21,22].

In this work, we introduce a neural network designed to efficiently handle the extensive datasets produced by surveys while minimizing the preprocessing required for automatic classification of variable stars. Our method eliminates the need for feature computation, scales seamlessly to large datasets, and requires only the known periods and corresponding light curves. We demonstrate our approach through experimental analysis using the ASAS-SN dataset.

The rest of the paper is structured as follows. Section 2 describes the data used for training and testing, the classification models, and the preprocessing steps necessary for CNN implementation. Section 3 details the architecture of the proposed neural network. Section 4 presents the results and discusses their implications. Section 5 concludes the paper.

## 2. DATASET AND PREPROCESSING

### 2.1 ALL-SKY AUTOMATED SURVEY FOR SUPERNOVAE (ASAS-SN)

We obtained the time-series data from the ASAS-SN survey. Since 2014, ASAS-SN has conducted sky surveys in the V-band with a limiting magnitude of $V \lesssim 17$ mag and a cadence of approximately 2–3 days, utilizing 8 telescopes in Chile and Hawaii. In 2018, the survey transitioned to the g-band and expanded its instrumentation to 20 cameras on 5 mounts, adding new units in South Africa, Texas, and Chile. All ASAS-SN telescopes are operated by the Las Cumbres Observatory. Compared to the V-band data, the g-band data offers improved depth ($g \leq 18.5$ mag), higher cadence ($\leq 24$ hours in the g-band versus 2–3 days in the V-band), and reduced diurnal aliasing due to the survey's geographically distributed units.

We used g-band data from ASAS-SN, focusing on variables with high-quality data. ASAS-SN provides a classification probability for each variable star, and we selected those with a classification probability of $\geq 0.99$. Each variable star's light curve was reviewed, and data points with inaccurate magnitude measurements, where the magnitude error exceeded 99%, were removed. Additionally, light curves with fewer than 50 total data points were excluded to ensure sufficient data for reliable analysis. Single-point outliers deviating by more than 3σ, likely caused by instrumental effects, were also removed.

To ensure stable training of the classification model, we selected only classes with an adequate number of distinct light curves (~ 400 or more). Due to the limited number of Cepheid light curves (405 in total), we restricted the other classes to a maximum of 5000 light curves to prevent any class from overwhelming influence in training. Rather than attempting to include all known variable star classes, as done with limited success in [12], we focused on a subset of six well-studied classes. These classes are Cepheids, δ Scuti, Mira, RR Lyrae, Eclipsing binaries, and Semi-regular variables. Our final dataset from ASAS-SN comprised a total of 21,344 light curves. A breakdown of the dataset and the number of light curves per class is provided in Table 1. Typical phase-folded light curves for each class are presented in Figure 1.

| Class | Representation | Number |
| --- | --- | --- |
| Cepheids | CEP | 405 |
| δ Scuti | DSCUT | 939 |
| Mira | M | 5,000 |
| RR Lyrae | RR | 5,000 |
| Eclipsing binaries | ECL | 5,000 |
| Semi-regular | SR | 5,000 |

Table 1. Total numbers of light curves per class in our dataset.

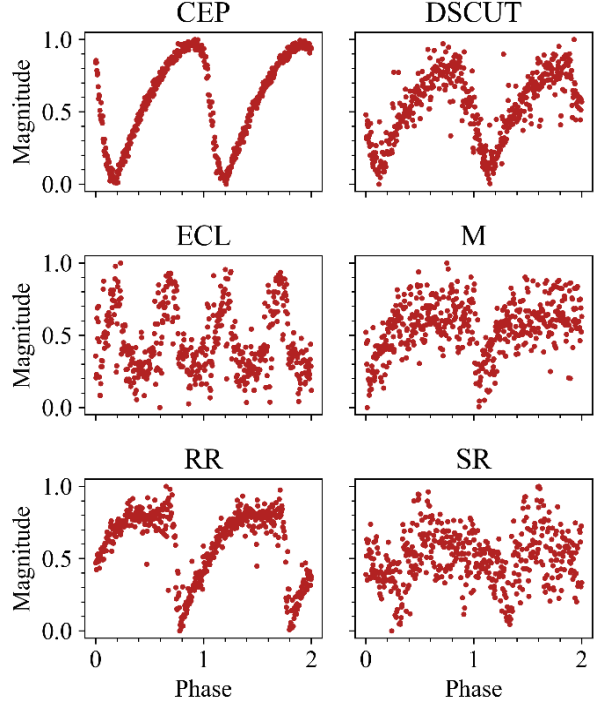

Figure 1. Illustrative examples of light curve profiles for variable stars. The titles indicate the corresponding star types. While most profiles are visibly distinct, some classes can be challenging to differentiate. Magnitudes are normalized for clarity.

## 2.2 INTERPOLATED PHASE-FOLDED LIGHT CURVES

The choice of a 1D CNN model is driven by the representation of light curves as one-dimensional sequences, where altering or reordering the data can result in the loss of crucial information. For periodic variability, as observed in our dataset with known periods, phase-folded light curves provide deeper insights compared to raw data. This is shown in Figure 2, which shows a Cepheid light curve from the ASAS-SN survey. The raw light curve in the left panel lacks a discernible pattern, while the phase-folded version in the right panel reveals a distinct sinusoidal pattern, emphasizing its periodic nature. Thus, phase-folded light curves offer more informative inputs, enhancing the model's ability to learn meaningful patterns.

We propose a 1D CNN model that uses phase-folded light curves as input. To ensure compatibility with the model, all input sequences were standardized to a uniform length of 512 points using interpolation. This process involves resampling each light curve to match a predefined number of points, ensuring consistency across the dataset while preserving the critical features of the light curves. Interpolation allows for the retention of essential temporal and amplitude information, avoiding the loss of variability patterns that are crucial for classification. By standardizing the input lengths, the model is better equipped to learn meaningful patterns from the phase-folded light curves, optimizing its performance while maintaining the integrity of the data.

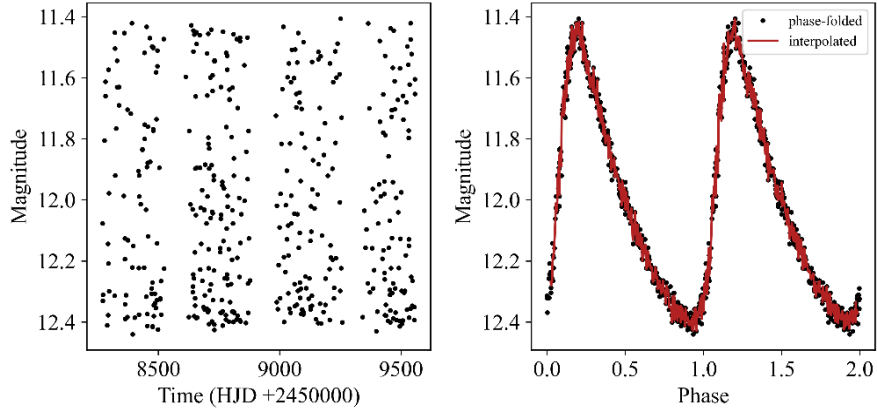

Figure 2. Light curve for a Cepheid from the ASAS-SN survey. The left panel shows the original light curve, while the right panel displays the phase-folded light curve with an estimated period of 4.36 days.

### 3. CONVOLUTIONAL NEURAL NETWORK

Convolutional neural networks (CNNs or ConvNets) are highly regarded as effective tools for pattern recognition across various domains of astronomical data [23–25]. They have been widely utilized in astronomy for tasks involving both classification and regression. Multiple studies have shown that deep learning frameworks, including CNNs, frequently surpass the performance of traditional machine learning algorithms.

CNN is a type of deep neural network specifically designed for extracting features and recognizing patterns, particularly in image data. It typically consists of an input layer, an output layer, and several hidden layers, such as convolutional, activation, pooling, fully connected, and normalization layers. Convolutional layers apply filters to the input data, generating feature maps that are passed to subsequent layers. Pooling layers, such as MaxPooling, reduce the dimensionality of the feature maps by selecting the maximum value within a given region, which helps improve generalization and speed up training. Average Pooling, on the other hand, computes the average value within the pooling window. Fully connected layers interpret the features extracted by previous layers, with each neuron connected to all activations from the preceding layer. Normalization layers stabilize the learning process by adjusting activations, helping to prevent overfitting.

In this work, we utilized the CNN architecture illustrated in Figure 3, based on the design described in [26,27]. The model consists of four convolutional blocks followed by two dense layers, with the output layer employing a softmax activation function for classification. Each convolutional block includes a 1D convolutional layer with ReLU activation, a batch normalization layer placed after the convolutional layer, and a 1D MaxPooling layer. The Conv1D layers have a kernel size of 3 and use "same" padding, while the MaxPooling1D layers have a pool size of 2. The convolutional layers are configured sequentially with filter sizes of 128, 64, 32, and 32, enabling the model to progressively extract features at different levels of abstraction. A dropout layer is applied before the final convolutional block to prevent overfitting by randomly deactivating neurons during training. The dense layers are designed to further process the extracted features, with the hidden layer containing 128 neurons activated by ReLU and a dropout layer applied afterward. The output layer contains 6 neurons with a softmax activation function to handle multi-class classification tasks. This architecture includes a total of 166,662 trainable parameters. The CNN was implemented and trained using TensorFlow 2.1 [28], with the Adam optimizer [29] set to a learning rate of 0.0001 and a batch size of 16. Categorical cross-entropy was used as the loss function. The model was trained over 100 epochs, with each epoch taking about 4 seconds.

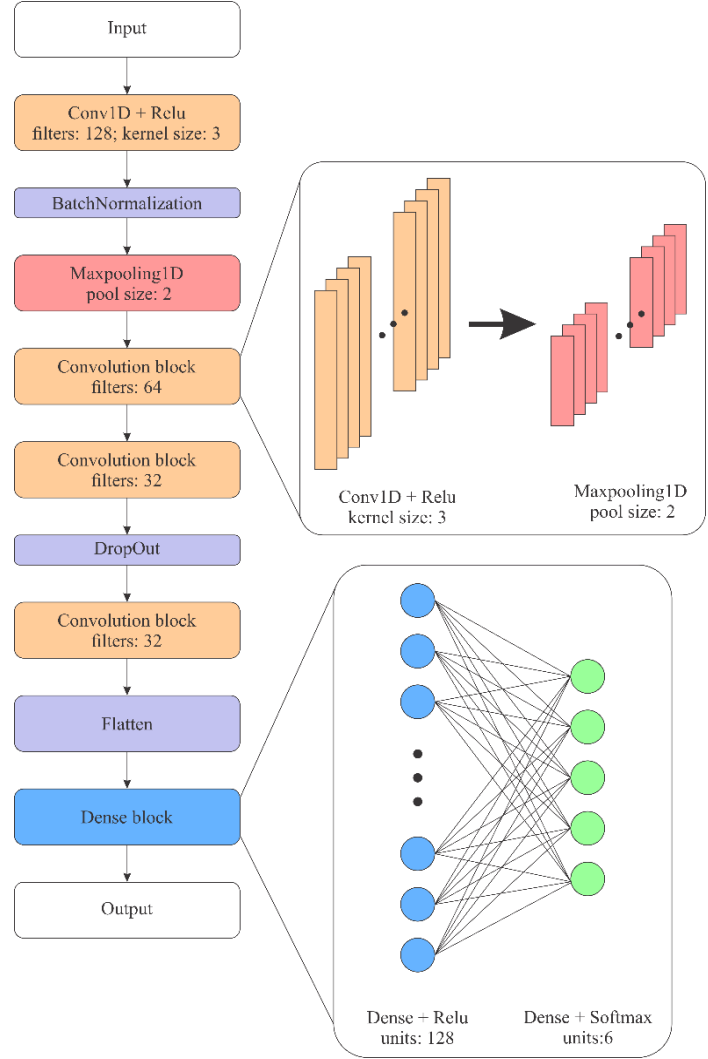

Figure 3. Scheme of the optical integrated system of light therapeutic influence.

## 4. RESULTS AND DISSCUSION

The objective of our experiments was to assess the classification accuracy of the model detailed in Section 3. The results were obtained by dividing the labeled dataset into an 80% training set and a 20% testing set. The training process was conducted using a single NVIDIA GeForce RTX 4070 Ti GPU with 12 GB of video memory.

Table 2 provides four key metrics: accuracy, precision, recall, and F1 score. Each of these metrics is defined for an individual class as follows:

$$Accuracy_i = \frac{TP_i + TN_i}{TP_i + TN_i + FP_i + FN_i} \tag{1}$$

$$Precision_i = \frac{TP_i}{TP_i + FP_i} \tag{2}$$

$$Recall_i = \frac{TP_i}{TP_i + FN_i} \tag{3}$$

$$F1\ score_i = 2 * \frac{Precision_i * Recall_i}{Precision_i + Recall_i}, \tag{4}$$

where $TP_i$ denotes the number of true positives, $FP_i$ represents the number of false positives, and $FN_i$ indicates the number of false negatives for a specific class $i$. Despite the substantial imbalance present in the labeled dataset, all classes are treated as equally important; therefore, we calculated averaged scores as:

$$Accuracy_{average} = \frac{1}{n_{cl}} \sum_{i=1}^{n_{cl}} Accuracy_i \qquad (5)$$

$$Precision_{average} = \frac{1}{n_{cl}} \sum_{i=1}^{n_{cl}} Precision_i \qquad (6)$$

$$Recall_{average} = \frac{1}{n_{cl}} \sum_{i=1}^{n_{cl}} Recall_i \qquad (7)$$

$$F1\ score_{average} = \frac{1}{n_{cl}} \sum_{i=1}^{n_{cl}} F1\ score_i, \qquad (8)$$

where $n_{cl}$ denotes the total number of classes.

| Class | Accuracy, % | Precision | Recall | F1 score |
|---|---|---|---|---|
| CEP | 84.4 ± 4.72 | 0.73 ± 0.06 | 0.84 ± 0.05 | 0.75 ± 0.04 |
| DSCUT | 90.0 ± 3.53 | 0.81 ± 0.01 | 0.90 ± 0.04 | 0.73 ± 0.07 |
| ECL | 98.6 ± 0.25 | 0.99 ± 0.01 | 0.99 ± 0.01 | 0.99 ± 0.01 |
| M | 84.3 ± 4.80 | 0.92 ± 0.01 | 0.80 ± 0.05 | 0.86 ± 0.03 |
| RR | 92.3 ± 2.31 | 0.91 ± 0.02 | 0.92 ± 0.02 | 0.92 ± 0.01 |
| SR | 90.9 ± 1.45 | 0.93 ± 0.02 | 0.91 ± 0.02 | 0.92 ± 0.01 |
| **Average** | **90.1 ± 2.84** | **0.88 ± 0.02** | **0.89 ± 0.03** | **0.86 ± 0.02** |

Table 2. Classification metrics (mean ± standard deviation) of our model. The values and errors were obtained from 5 runs using the k-fold cross-validation method.

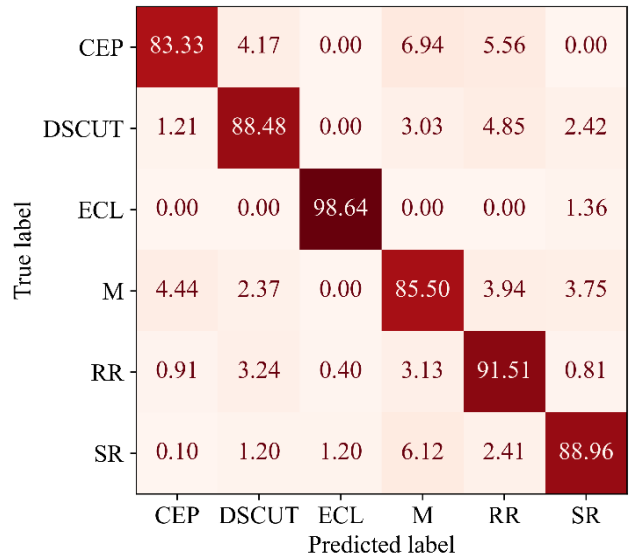

Figure 4. Classification results of our CNN model from one of the runs.

The classification model demonstrates strong performance across various star classes, with an average accuracy of 90.1%, and robust precision (0.88), recall (0.89), and F1 score (0.86). The ECL class shows particularly high performance, with nearly perfect metrics (accuracy of 98.6%, precision and recall of 0.99), indicating that the model can reliably distinguish ECL stars. The RR and SR classes also perform well, with accuracy scores of 92.3% and 90.9%, respectively, along with balanced precision, recall, and F1 scores, suggesting consistent and accurate classification. The DSCUT class achieves a high recall (0.90), but its lower precision (0.81) and F1 score (0.73) point to occasional misclassifications. The M class shows high precision (0.92), though its slightly lower recall (0.80) results in a higher rate of missed stars, as indicated by the F1 score of 0.86. The CEP class exhibits a notable difference between precision (0.73) and recall (0.84), where the lower precision suggests a higher incidence of false positives. Despite this, the model's relatively high recall indicates its effectiveness in identifying most CEP stars.

## 5. CONCLUSIONS

In this study, we present a CNN-based classifier for identifying the type of variable stars using light curves. Trained and evaluated on the ASAS-SN dataset, our model achieves an average accuracy of 90% and an F1 score of 0.86 across six well-known classes of variable stars. The model effectively handles the diverse shapes and sampling cadences of light curves, offering a robust and automated solution for star classification. This approach removes the need for manual feature computation, is scalable to large datasets, and requires only the known periods alongside the light curves, making it highly efficient for handling extensive astronomical data.

The CNN-based method presented in this article provides a data-driven solution for classifying variable stars with high accuracy and sensitivity. Its ability to efficiently process large volumes of data from current and future sky surveys positions it as a valuable tool for astronomical research. This automated method not only enhances the precision of star classification but also facilitates the analysis of vast datasets in a more streamlined manner. Overall, the model shows great promise for advancing the field of variable star classification and contributing to the ongoing study of stellar behavior across the sky.

## 6. ACKNOWLEDGMENTS

We would like to express deepest appreciation to the Research Institute of Experimental and Theoretical Physics at alFarabi Kazakh National University for their invaluable support in providing computing resources of the Department of Physics and Technology for this study.

## 7. FUNDING

The research received funding from the Ministry of Science and Higher Education of the Republic of Kazakhstan, under grant number AP19674715.

*Authors: Master's student, Junior Researcher of Department of Electronics and Astrophysics, Akhmetali A., Al-Farabi Kazakh National University, 71 al-Farabi Avenue, 050040, Almaty, Kazakhstan, E-mail: akhmetali_almat@kaznu.edu.kz, Master's student, Junior Researcher of Department of Electronics and Astrophysics, Zhunuskanov A., Al-Farabi Kazakh National University, 71 al-Farabi Avenue, 050040, Almaty, Kazakhstan, Master of Natural Sciences, Senior Lecturer of Department of Solid State Physics and Nonlinear Physics, Namazbayev T., Al-Farabi Kazakh National University, 71 al-Farabi Avenue, 050040, Almaty, Kazakhstan, Student, Department of Electronics and Astrophysics, Zaidyn M., Al-Farabi Kazakh National University, 71 al-Farabi Avenue, 050040, Almaty, Kazakhstan, Master's student, Junior Researcher of Department of Electronics and Astrophysics, Sakan A., Al-Farabi Kazakh National University, 71 al-Farabi Avenue, 050040, Almaty, Kazakhstan, PhD student, Senior Lecturer of Department of Solid State Physics and Nonlinear Physics, Turlykozhayeva D.A., Al-Farabi Kazakh National University, 71 al-Farabi Avenue, 050040, Almaty, Kazakhstan, PhD student, Senior Lecturer of Department of Solid State Physics and Nonlinear Physics, Ussipov N.M., Al-Farabi Kazakh National University, 71 al-Farabi Avenue, 050040, Almaty, Kazakhstan.*